\documentclass[eqsecnum,twocolumn,aps,epsf]{revtex4}
\usepackage{graphicx}
\usepackage{color}
\begin{document}
\draft

\title{Muon-spin-relaxation and magnetic-susceptibility studies of effects of the magnetic impurity Ni on the Cu-spin dynamics and superconductivity in La$_{2-x}$Sr$_x$Cu$_{1-y}$Ni$_y$O$_4$ with $x=0.13$}

\author{T. Adachi, S. Yairi, Y. Koike}

\address{Department of Applied Physics, Graduate School of Engineering, Tohoku University,\\Aoba-yama 05, Aoba-ku, Sendai 980-8579, Japan}

\author{I. Watanabe}

\address{Muon Science Laboratory, RIKEN (The Institute of Physical and Chemical Research), 2-1 Hirosawa, Wako 351-0198, Japan}

\author{K. Nagamine}

\address{Meson Science Laboratory, Institute of Materials Structure Science, High Energy Accelerator Research Organization, 1-1 Oho, Tsukuba 305-0801, Japan}
\date{\today}

\begin{abstract}

Effects of the magnetic impurity Ni on the Cu-spin dynamics and superconductivity have been studied in La$_{2-x}$Sr$_x$Cu$_{1-y}$Ni$_y$O$_4$ with $x = 0.13$ changing $y$ finely up to 0.10. 
Compared with the case of the nonmagnetic impurity Zn, it has been found from the muon-spin-relaxation measurements that a large amount of Ni is required to stabilize a magnetic order of Cu spins. 
However, the evolution toward the stabilization of the magnetic order with increasing impurity concentration is qualitatively similar to each other. 
The area of the non-superconducting and slowly fluctuating or static region of Cu spins around Ni has been found to be smaller than that around Zn, suggesting that the pinning of rather long-ranged dynamical spin correlation such as the so-called dynamical stripe by Ni is weaker than that by Zn. 
This may be the reason why Zn destroys the superconductivity in the hole-doped high-$T_c$ cuprates more markedly than Ni. 

\end{abstract}
\vspace*{2em}
\maketitle
\newpage

%*****************************************************************************************
%\section{Introduction}\label{intro}
%*****************************************************************************************
In the history of the study of high-$T_{\rm c}$ superconductors, the impurity effect on the Cu-spin dynamics and superconductivity is one of central subjects. 
Especially, both nonmagnetic Zn$^{2+}$ (the spin quantum number $S=0$) and magnetic Ni$^{2+}$ ($S=1$) have been good candidates for the study of the impurity effect. 
It is widely recognized that local magnetic moments due to Cu spins are induced by Zn in the underdoped regime~\cite{mahajan} and that the superconductivity is strongly suppressed by Zn.~\cite{maeno} 
On the other hand, both the induction of local magnetic moments and the suppression of superconductivity by Ni are weaker than those by Zn.~\cite{ishida} 
In this regard, the earlier NMR and NQR measurements have revealed that Zn acts as a strong scatterer of holes while Ni acts as a weak scatterer~\cite{kitaoka}, though the reason has not yet been clarified. 

The impurity effect on the dynamical stripe correlations of spins and holes suggested in the La-214 system~\cite{nature} has attracted great interest. 
From the experimental~\cite{koike3,adachi} and theoretical~\cite{smith} works, the dynamical stripe correlations have been proposed to be pinned and stabilized by the nonmagnetic Zn. 
The formation of such a static stripe order is considered to lead to the suppression of superconductivity, which is typically observed around the hole concentration of 1/8 per Cu (1/8 anomaly). 

Recently, we have investigated effects of Zn on the Cu-spin dynamics and superconductivity from the muon-spin-relaxation ($\mu$SR) and magnetic-susceptibility $\chi$ measurements in La$_{2-x}$Sr$_x$Cu$_{1-y}$Zn$_y$O$_4$ around $x=0.115$ changing $y$ finely up to 0.10.~\cite{nabe,nabe2,adachi2,adachi3} 
It has been found that the volume fraction of the superconducting region estimated from $\chi$ decreases rapidly through the slight doping of Zn and that its $y$ dependence corresponds to the $y$ dependence of the volume fraction of the fast fluctuating region of Cu spins estimated from the $\mu$SR results. 
It has been concluded that the non-superconducting region induced around Zn corresponds to the region where the Cu-spin fluctuations slow down, in other words, both the slowly fluctuating region of Cu spins and the magnetically ordered one compete with the superconductivity. 

In this paper, we investigate effects of Ni on the Cu-spin dynamics and superconductivity from the zero-field (ZF) $\mu$SR and $\chi$ measurements in the Ni-substituted La$_{2-x}$Sr$_x$Cu$_{1-y}$Ni$_y$O$_4$ with $x=0.13$ changing $y$ finely up to 0.10.~\cite{adachi4} 
Our aim is to clarify the difference of the impurity effect between the nonmagnetic Zn and magnetic Ni. 
In particular, we try to clarify whether the above-mentioned model proposed in the Zn-substituted case is also applicable to the Ni-substituted case or not. 

%*****************************************************************************************
%\section{Experimental details}
%*****************************************************************************************
\begin{figure}[tbp]
\begin{center}
\includegraphics[width=0.85\linewidth]{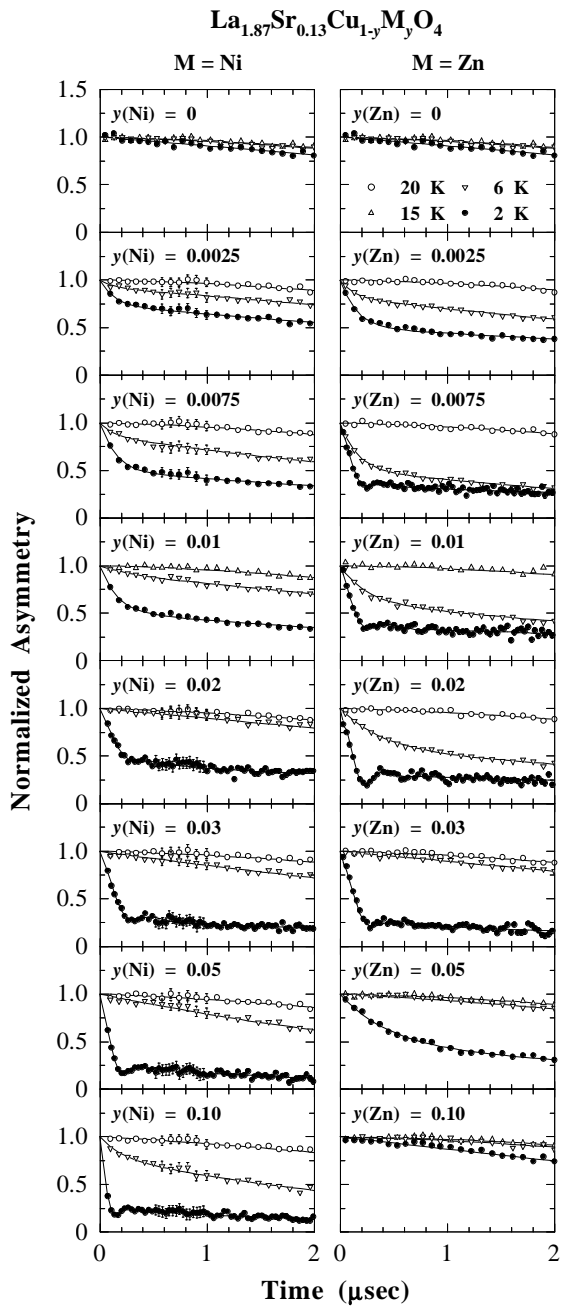}
\end{center}
\caption{ZF-$\mu$SR time spectra for typical $y$ values of La$_{2-x}$Sr$_x$Cu$_{1-y}$M$_y$O$_4$ (M = Ni, Zn) with $x=0.13$ at various temperatures down to 2 K. Solid lines indicate the best-fit results using $A(t) = A_0 e^{-\lambda_0t}G_Z(\Delta,t) + A_1 e^{-\lambda_1t} + A_2 e^{-\lambda_2t}{\rm cos}(\omega t + \phi)$.}  
\label{spec} 
\end{figure}

Polycrystalline samples of La$_{2-x}$Sr$_x$Cu$_{1-y}$Ni$_y$O$_4$ with $x=0.13$ and $y=0 - 0.10$ were prepared by a conventional solid-state reaction method. 
The detailed procedures were almost the same as those in the Zn-substituted case.~\cite{adachi3} 
All of the samples were checked by the powder X-ray diffraction measurements to be single phase. 
The electrical resistivity was also measured to check the quality of the samples. 
The ZF-$\mu$SR measurements were performed at the RIKEN-RAL Muon Facility at the Rutherford-Appleton Laboratory in the UK. 
The $\mu$SR time spectrum, namely, the time evolution of the asymmetry parameter $A(t)$ was measured down to 2 K to investigate the Cu-spin dynamics. 
The $\chi$ measurements were carried out down to 2 K using a standard SQUID magnetometer in a magnetic field of 10 Oe on field cooling, in order to evaluate the volume fraction of the superconducting region.~\cite{adachi3} 

%*****************************************************************************************
%\section{Results and Discussion}
%*****************************************************************************************
Figure \ref{spec} displays the typical ZF-$\mu$SR time spectra of La$_{2-x}$Sr$_x$Cu$_{1-y}$Ni$_y$O$_4$ with $x=0.13$ and $y=0 - 0.10$, together with the Zn-substituted data reported in our previous papers.~\cite{nabe,nabe2,adachi2,adachi3}
At a high temperature of 15 K or 20 K, all the spectra show little depolarization in the early time region from 0 to 2 $\mu$sec, indicating almost no influence of Cu spins. 
In this case, only a Gaussian-like depolarization due to randomly oriented nuclear spins is observed. 
Focusing on the spectra at 2 K, the Gaussian-like depolarization is still observed in the non-substituted sample of $y=0$, indicating a fast fluctuating state of Cu spins beyond the $\mu$SR time window (10$^{-6} - 10^{-11}$ sec). 
In the Ni-substituted case, the muon-spin depolarization becomes fast with increasing $y$(Ni), indicating slowing down of the Cu-spin fluctuations. 
Muon-spin precession corresponding to the formation of a coherent magnetic order is observed for $y$(Ni) $\ge0.02$ and gradually develops up to $y$(Ni) $=0.10$. 
In the Zn-substituted case, the muon-spin precession starts to be observed at $y$(Zn) $=0.0075$, and the magnetic order develops most at $y$(Zn) $=0.02$. 
The muon-spin precession disappears for $y$(Zn) $>0.03$ and almost no fast depolarization of muon spins is observed at $y$(Zn) $=0.10$, meaning that the static magnetic order is destroyed and that Cu spins are fluctuating fast beyond the $\mu$SR time window. 
Compared with the Zn-substituted spectra, two important features are pointed out in the Ni-substituted spectra. 
One is that the evolution of the spectra with increasing $y$ is qualitatively similar to each other for small $y$ values, but the $y$ value where the muon-spin precession starts to be observed is larger in the Ni-case than in the Zn-case. 
The other is that a large amount of Ni does not destroy the magnetic order, whereas a large amount of Zn destroys it. 

The time spectra are analyzed with the following three-component function: $A(t) = A_0 e^{-\lambda_0t}G_Z(\Delta,t) + A_1 e^{-\lambda_1t} + A_2 e^{-\lambda_2t}{\rm cos}(\omega t + \phi)$. 
The first term represents the slowly depolarizing component in a region where Cu spins are fluctuating fast beyond the $\mu$SR time window ($A_0$ region). 
The second term represents the fast depolarizing component in a region where the Cu-spin fluctuations slow down and/or an incoherent magnetic order is formed ($A_1$ region). 
The third term represents the muon-spin precession in a region where a coherent static magnetic order of Cu spins is formed ($A_2$ region). 
The $A_0$, $A_1$, $A_2$ and $\lambda_0$, $\lambda_1$ are the initial asymmetries and depolarization rates of each component, respectively. 
The $G_Z(\Delta,t)$ is the static Kubo-Toyabe function describing the distribution of the nuclear-dipole field at the muon site.~\cite{uemura} 
The $\lambda_2$, $\omega$ and $\phi$ are the damping rate, frequency and phase of the muon-spin precession, respectively. 
The time spectra are well fitted with the above function, as shown by solid lines in Fig. \ref{spec}. 

\begin{figure}[tbp]
\begin{center}
\includegraphics[width=0.79\linewidth]{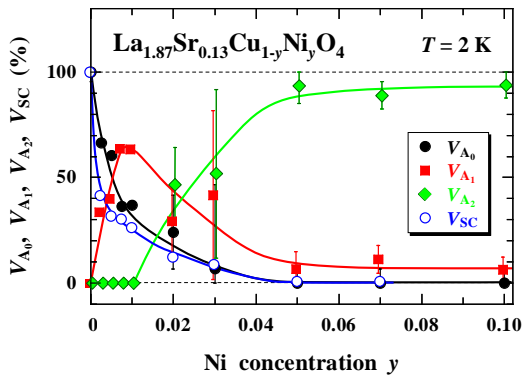}
\end{center}
\caption{(color) Dependence on $y$ of $V_{\rm A_0}$ (black circles), $V_{\rm A_1}$ (red squares) and $V_{\rm A_2}$ (green diamonds) estimated from the $\mu$SR measurements in La$_{2-x}$Sr$_x$Cu$_{1-y}$Ni$_y$O$_4$ with $x=0.13$. Error bars of each marks are derived from the errors of fitting using the function mentioned in the text. Dependence on $y$ of $V_{\rm SC}$ (blue circles) estimated from $\chi$ is also plotted. Solid lines are to guide the reader's eye.}  
\label{vol-fra} 
\end{figure}

Using the best-fit values of $A_0$, $A_1$ and $A_2$, we can estimate volume fractions $V_{\rm A_0}$, $V_{\rm A_1}$, $V_{\rm A_2}$ of the $A_0$, $A_1$, $A_2$ regions, respectively.~\cite{adachi2,adachi3} 
From the $\chi$ measurements, the volume fraction $V_{\rm SC}$ of the superconducting region can be estimated.~\cite{adachi2,adachi3} 
Figure \ref{vol-fra} displays the Ni-concentration dependence of $V_{\rm A_0}$, $V_{\rm A_1}$, $V_{\rm A_2}$ and $V_{\rm SC}$ at 2 K in La$_{2-x}$Sr$_x$Cu$_{1-y}$Ni$_y$O$_4$ with $x=0.13$. 
The $V_{\rm A_0}$ is 100 \% at $y=0$, indicating that all the Cu spins are fluctuating at high frequencies. 
With increasing $y$, $V_{\rm A_0}$ rapidly decreases and $V_{\rm A_1}$ increases, indicating slowing down of the Cu-spin fluctuations and/or the formation of an incoherent magnetic order. 
In fact, $V_{\rm A_2}$ increases instead of $V_{\rm A_1}$ for $y>0.01$. 
At $y=0.05$, $V_{\rm A_0}$ becomes zero and almost all the Cu spins become static and are magnetically ordered. 
The magnetically ordered state seems to be maintained up to $y=0.10$. 
It is found that the $y$ dependence of $V_{\rm SC}$ is in rough agreement with that of $V_{\rm A_0}$, which is similar to the Zn-case.~\cite{adachi2,adachi3} 
These results strongly suggest that, even in the Ni-case, the superconductivity is realized in the region where Cu spins fluctuate fast beyond the $\mu$SR time window. 
In other words, it appears that Cu spins in a non-superconducting region around Ni exhibit slowing down of the fluctuations or form an incoherent or coherent static order. 

\begin{figure}[tbp]
\begin{center}
\includegraphics[width=0.7\linewidth]{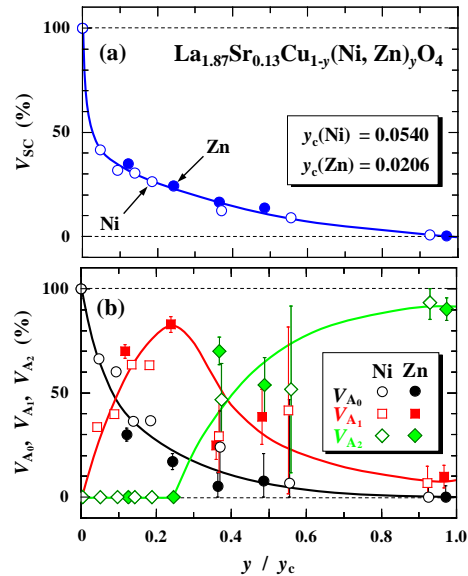}
\caption{(color) Dependence on $y$, normalized by the value of $y_c$ where $V_{SC}$ is extrapolated to be zero, of (a) $V_{SC}$ and (b) $V_{A_0}$ (circles), $V_{A_1}$ (squares) and $V_{A_2}$ (diamonds) at 2 K for La$_{2-x}$Sr$_x$Cu$_{1-y}$(Ni, Zn)$_y$O$_4$ with $x=0.13$. Solid lines are to guide the reader's eye.}
\label{normalize}
\end{center}
\end{figure}

As seen in Fig. \ref{vol-fra}, the change of $V_{\rm A_0}$, $V_{\rm A_1}$, $V_{\rm A_2}$ and $V_{\rm SC}$ with increasing $y$ for small $y$ values in the Ni-case is analogous to that in the Zn-case.~\cite{adachi2,adachi3} 
Figure \ref{normalize}(a) shows the dependence of $V_{\rm SC}$ on $y$, normalized by the value of $y_c$ where $V_{\rm SC}$ is extrapolated to be zero in the Ni- ($y_c$(Ni) $= 0.0540$) and Zn- ($y_c$(Zn) $= 0.0206$) cases. 
Apparently, $V_{\rm SC}$'s in the Ni- and Zn-cases are located on the same line. 
Furthermore, as shown in Fig. \ref{normalize}(b), $V_{\rm A_0}$, $V_{\rm A_1}$ and $V_{\rm A_2}$ in the Ni-case are also in rough correspondence to those in the Zn-case, respectively. 
These give an important information that the evolution toward the formation of the magnetic order with increasing impurity concentration is similar between the Ni- and Zn-cases. 
That is, changes of the hole and Cu-spin states with increasing impurity concentration are independent of the magnetism of impurities for $y<y_c$, though the value of $y_c$ depends on the impurity. 

\begin{figure}[tbp]
\begin{center}
\includegraphics[width=0.9\linewidth]{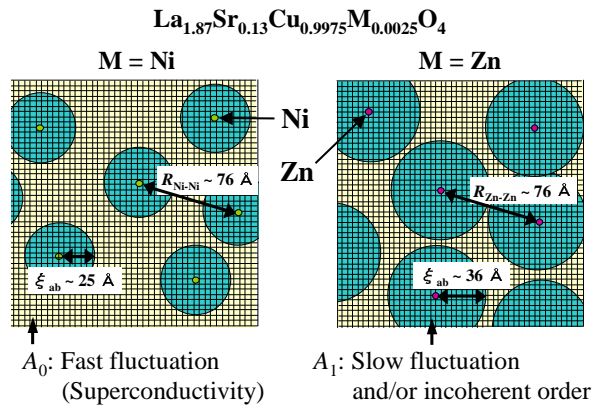}
\end{center}
\caption{(color) Schematic pictures of the spatial distribution of the different Cu-spin states in the CuO$_2$ plane, namely, the $A_0$ and $A_1$ regions in La$_{2-x}$Sr$_x$Cu$_{1-y}$M$_y$O$_4$ (M = Ni, Zn) with $x=0.13$ and $y=0.0025$. Each crossing point of the grid pattern represents the Cu site. Ni or Zn atoms are randomly distributed in the CuO$_2$ plane. The mean distance of Zn-Zn or Ni-Ni is $\sim 76 {\rm \AA}$. The size of circles is calculated from the ratio of $V_{\rm A_0}$: $V_{\rm A_1}$ in the Ni- and Zn-substituted cases, respectively.}  
\label{picture} 
\end{figure}

Figure \ref{picture} displays schematic pictures of the spatial distribution of the $A_0$ and $A_1$ regions for $y=0.0025$ in La$_{2-x}$Sr$_x$Cu$_{1-y}$(Ni, Zn)$_y$O$_4$ with $x=0.13$, on the assumption that Cu-spin fluctuations slow down {\it around} each Ni and Zn. 
The size of circles is calculated from the ratio of $V_{\rm A_0}$: $V_{\rm A_1}$ and $\xi_{ab}$ is the radius of the $A_1$ region, namely, a slowly fluctuating and/or incoherent magnetically ordered region of Cu spins. 
The value of $\xi_{ab}$ is estimated as $\sim 25 {\rm \AA}$ in the Ni-case and $\sim 36 {\rm \AA}$ in the Zn-case. 
It is apparent that the $A_1$ region is smaller around Ni than around Zn. 
Accordingly, $A_1$ regions cover the almost whole CuO$_2$ plane in the Zn-case, while $A_0$ regions corresponding to the fast fluctuation of Cu spins still remain extensively in the Ni-case. 
These indicate that, through the 0.25 \% substitution of impurities, $V_{\rm SC}$ in correspondence to $V_{\rm A_0}$ rather survives in the Ni-case while it is strongly diminished in the Zn-case. 

In our previous papers,~\cite{adachi2,adachi3} the Cu-spin state of the non-superconducting region around Zn has been discussed in terms of the pinning of the dynamical spin correlation such as the dynamical stripe. 
The $\xi_{ab}$ is the largest at $x=0.115$ in La$_{2-x}$Sr$_x$Cu$_{1-y}$Zn$_y$O$_4$, which is in accordance with the results that the correlation length of the dynamical spin correlation estimated from the inelastic neutron scattering measurements is the largest around $x=0.115$.~\cite{yamada} 
Therefore, it has been concluded that the dynamical spin correlation such as the dynamical stripe is pinned and stabilized by Zn. 
Based on this conclusion, it is likely that the dynamical spin correlation such as the dynamical stripe is pinned and stabilized by Ni as well. 
The reason why the pinned $A_1$ region around Ni is smaller than that around Zn may be explained in terms of the stripe model as follows. 
In the case of the dynamical stripe correlations, two cases of the pinning may be supposed; one is that Ni or Zn is located at a charge (hole) stripe and the other is at a spin stripe. 
Considering the exchange interaction between Cu spins $J$ and the transfer integral of holes $t$, the difference of the energy loss between two cases may be smaller in the Ni-case than in the Zn-case because of the difference in $S$ between Ni$^{2+}$ ($S=1$) and Zn$^{2+}$ ($S=0$). 
This means that Zn pins the dynamical stripe correlations more effective than Ni. 
Supposing the dynamical stripe correlations exist in a wide range of hole concentration where the superconductivity appears,~\cite{yamada} this may be the reason why Zn destroys the superconductivity in the hole-doped high-$T_c$ cuprates more markedly than Ni.~\cite{maeno} 
Moreover, the fact that Zn acts as a strong scatterer of holes~\cite{kitaoka} may be explained as the strong pinning of the dynamical stripe by Zn. 

Finally, we discuss the Cu-spin state for $y$(Ni) $>0.05$. 
As shown in Fig. \ref{vol-fra}, $V_{\rm A_2}$ is above 90 \% even at $y$(Ni) $=0.10$, meaning that the coherent static magnetic order is maintained up to $y$(Ni) $=0.10$. 
In the Zn-case, on the contrary, the magnetic order is destroyed at $y$(Zn) $=0.10$ owing to the spin-dilution effect of the nonmagnetic Zn$^{2+}$.~\cite{nabe,nabe2,adachi2,adachi3} 
As for the internal field $H_{\rm int}$ at the muon site, calculated from $\omega$ as $H_{\rm int} = \omega / \gamma_{\mu}$ ($\gamma_{\mu}$: the gyromagnetic ratio of the muon spin), it is $\sim 150$ gauss and $\sim 130$ gauss at $y$(Ni) $=0.03$ and $y$(Zn) $=0.03$, respectively. 
These suggest that a similar magnetic order is formed at $y\sim 0.03$ in both Ni- and Zn-cases. 
This is reasonably understood in terms of the above-mentioned stripe-pinning model. 
With increasing $y$(Ni), $H_{\rm int}$ increases gradually and reaches $\sim 270$ gauss at $y$(Ni) $= 0.10$. 
The inclusion of Ni$^{2+}$ ($S=1$) in the Cu$^{2+}$ ($S=1/2$) sea may cause an increase of the mean value of $S$, leading to the increase of $H_{\rm int}$ up to $\sim 270$ gauss. 
As for $\lambda_2$, it is almost similar to each other in the Ni- and Zn-cases at $y\sim0.03$. 
However, $\lambda_2$ increases with increasing $y$(Ni) for $y$(Ni) $> 0.03$, suggesting that a large amount of Ni tends to depress the coherency of the magnetic order. 
Therefore, the magnetic order for $y$(Ni) $\sim 0.10$ may be regarded as a precursory state toward the spin-glass state, which is reminiscent of that observed from the $\chi$ measurements in La$_{2-x}$Sr$_x$Cu$_{1-y}$Ni$_y$O$_4$ with $x=0.08$ and $y\ge0.2$.~\cite{machi} 
Based on the stripe model, it is possible that the less coherent magnetically ordered state for $y$(Ni) $\sim 0.10$ is a slightly disordered stripe state. 

%*****************************************************************************************
%\section{Conclusion}
%*****************************************************************************************
In conclusion, compared with the case of the nonmagnetic impurity Zn, a large amount of Ni is required to stabilize a magnetic order of Cu spins. 
However, the evolution toward the stabilization of the magnetic order is qualitatively similar to each other. 
The area of the non-superconducting and slowly fluctuating or static region of Cu spins around Ni is smaller than that around Zn, suggesting that the pinning of the dynamical spin correlation such as the dynamical stripe by Ni is weaker than that by Zn. 
This may be the reason why Zn destroys the superconductivity in the hole-doped high-$T_c$ cuprates more markedly than Ni. 
While the magnetic order is destroyed through the 10 \% substitution of Zn, it is maintained through the 10 \% substitution of Ni. 
A slightly disordered stripe state may be realized for $y$(Ni) $\sim 0.10$. 

%*****************************************************************************************
%\section*{Acknowledgments}
%*****************************************************************************************
This work was supported by a Grant-in-Aid for Scientific Research from the Ministry of Education, Science, Sports, Culture and Technology, Japan, and also by CREST of Japan Science and Technology Corporation.


\begin{references}
\bibitem{mahajan}
A.V. Mahajan {\it et al.}, Phys. Rev. Lett. {\bf 72}, 3100 (1994).

\bibitem{maeno}
Y. Maeno {\it et al.}, Physica C {\bf 153-155}, 1105 (1988).

\bibitem{ishida}
K. Ishida {\it et al.}, J. Phys. Soc. Jpn. {\bf 62}, 2803 (1993).

\bibitem{kitaoka}
Y. Kitaoka {\it et al.}, J. Phys. Soc. Jpn. {\bf 63}, 2052 (1994).

\bibitem{nature}
J.M. Tranquada {\it et al.}, Nature (London) {\bf 375}, 561 (1995).

\bibitem{koike3}
Y. Koike {\it et al.}, Physica C {\bf 282-287}, 1233 (1997).
 
\bibitem{adachi}
T. Adachi {\it et al.}, J. Low Temp. Phys. {\bf 117}, 1151 (1999).

\bibitem{smith}
C.M. Smith {\it et al.}, Phys. Rev. Lett. {\bf 87}, 177010 (2001).

\bibitem{nabe}
I. Watanabe {\it et al.}, J. Phys. Chem. Solids {\bf 63}, 1093 (2002).

\bibitem{nabe2}
I. Watanabe {\it et al.}, Phys. Rev. B {\bf 65}, 180516(R) (2002).

\bibitem{adachi2}
T. Adachi {\it et al.}, J. Low Temp. Phys. {\bf 131}, 843 (2003).

\bibitem{adachi3}
T. Adachi {\it et al.}, cond-mat/0306233.

\bibitem{adachi4}
T. Adachi {\it et al.}, to be published in Physica C. 

\bibitem{uemura}
Y.J. Uemura {\it et al.}, Phys. Rev. B {\bf 31}, 546 (1985).

\bibitem{yamada}
K. Yamada {\it et al.}, Phys. Rev. B {\bf 57}, 6165 (1998).

\bibitem{machi}
T. Machi {\it et al.}, Physica C {\bf 388-389}, 233 (2003).

\end{references}
\end{document}